\documentclass[journal]{IEEEtran}

\usepackage{amsmath,amssymb,dblfloatfix}

\usepackage{booktabs,lipsum,makecell}
\usepackage{color,soul,xcolor,mathtools}
\usepackage{xcolor}
\usepackage[colorlinks = true,
            linkcolor = blue,
            urlcolor  = blue,
            citecolor = blue,
            anchorcolor = blue]{hyperref}
\newcommand{\MYhref}[3][blue]{\href{#2}{\color{#1}{#3}}}%
\usepackage{cite}

\usepackage[linesnumbered,ruled]{algorithm2e}
\usepackage{algpseudocode}
\usepackage{graphicx}

\usepackage{pifont}
\usepackage{multirow}
\usepackage{footnote}

\begin{document}
\title{SkipConvGAN: Monaural Speech Dereverberation using Generative Adversarial Networks via Complex Time-Frequency Masking}

\author{Vinay Kothapally,~\IEEEmembership{Student Member,~IEEE,} John H. L. Hansen,~\IEEEmembership{Fellow,~IEEE}
          
\thanks{V. Kothapally is with the Center for Robust Speech Systems, University of
	Texas at Dallas, Richardson, TX 75080 USA (e-mail: vinay.kothapally@utdallas.edu)}

\thanks{J. H. L. Hansen is with the Center for Robust Speech Systems, University of
	Texas at Dallas, Richardson, TX 75080 USA (e-mail: john.hansen@utdallas.edu)}}

\maketitle

\begin{abstract}
\label{sec:Abs}
With the advancements in deep learning approaches, the performance of speech enhancing systems in the presence of background noise have shown significant improvements. However, improving the system's robustness against reverberation is still a work in progress, as reverberation tends to cause loss of formant structure due to smearing effects in time and frequency. A wide range of deep learning-based systems either enhance the magnitude response and reuse the distorted phase or enhance complex spectrogram using a complex time-frequency mask. Though these approaches have demonstrated satisfactory performance, they do not directly address the lost formant structure caused by reverberation. We believe that retrieving the formant structure can help improve the efficiency of existing systems. In this study, we propose SkipConvGAN - an extension of our prior work SkipConvNet. The proposed system's generator network tries to estimate an efficient complex time-frequency mask, while the discriminator network aids in driving the generator to restore the lost formant structure. We evaluate the performance of our proposed system on simulated and real recordings of reverberant speech from the single-channel task of the  REVERB challenge corpus. The proposed system shows a consistent improvement across multiple room configurations over other deep learning-based generative adversarial frameworks.
\end{abstract}


%
\IEEEpeerreviewmaketitle

\section{Introduction}
\label{sec:Intro}
\IEEEPARstart{A}{s the sound} propagates in naturalistic environments such as conference rooms, lobbies or cafeterias, speech captured by a device is perceived as a distorted mixture of clean speech, intrusive background noise, and its own reflections from surrounding objects and surfaces. As a result, speech captured by distant microphones in such confined spaces faces two major challenges: (a) background noise: speech from multiple overlapping speakers, music, or other acoustical sounds picked up from the environment; (b) reverberation: reflection-induced self-distortion. Speech intelligibility is greatly reduced in either or both circumstances. These challenges are well-known in the research community and have been addressed through the use of a variety of signal processing and deep neural network (DNN) techniques. The terms ``speech denoising" and ``speech dereverberation" refer to systems developed to address them. These systems fall within the broader category of ``speech enhancement", which is the task of recovering a desired speaker's clean speech in a noisy and reverberant environment. Speech enhancing techniques for eliminating background noise or reducing the impact of reverberation from captured speech signals have been shown to be beneficial in a wide range of back-end speech applications, including automatic speech recognition (ASR) systems, speaker identification (SID), hearing aids for cochlear implants, teleconferencing systems, hands-free communication, and voice control for human-assisting devices.

The speech denoising aspect of enhancement is a widely used technique aimed at separating the desired audio signal from the intrusive noise \cite{denoising1}. Background noises of various types and levels, i.e., signal-to-noise ratio (SNR), have a direct impact on the performance of these systems. Background noises that are very similar to the desired speech that are produced in cafeteria or lunchroom with lower SNRs, make it extremely difficult task for the denoising systems as they share the frequency bands with the desired speech \cite{denoising4}. 

On the other hand, speech dereverberation aspect of enhancement refers to the task of suppressing convolutive effects of reflections induced into speech by the environment. The major factors that contribute to the number of reflections induced from the environment are the room size, damping properties of surfaces in the environment, and the distance of the capturing microphone from the origin of the desired speech. Lower signal-to-reverberation modulation energy ratio (SRMR) \cite{SRMR1,SRMR2} conditions, which correlate to stronger reflections, significantly reduce the performance of the dereverberation systems because the reflections cause smearing effects across time in the frequency bands with the desired speech. Researchers have introduced multiple techniques using single microphones over time, as well as techniques incorporating multiple microphones to leverage spatial information of desired speech, background noises, and reflections in addition to acoustical information. In this study, we focus exclusively on the speech dereverberation aspect of enhancement to improve the performance of systems using a single microphone.

One early attempt to reduce the effects of reverberation was to identify distinct peaks in the cepstrum of the speech signal and design a comb filter to cancel them out. However, it was discovered that this approach is inefficient for severe reverberant conditions \cite{oppenheim_comb}. Spectral subtraction, a widely used technique for speech denoising, was applied to dereverberation via statistical modeling of room impulse responses (RIR) using gaussian noise modulated with exponentially decaying functions \cite{labert}. The decay rate of the exponential function was used to govern the rate at which the reflections fade in a given environment viz., reverberation time. This technique has shown promising performance when extended for multi-microphone speech recordings. Likewise, the statistical estimation of room impulse response was used to compute an inverse filter capable of removing the effects of reverberation. This approach, however, could not produce satisfactory results for speech in realistic environments because it assumes the RIR function to be a minimum-phase function, which is frequently not satisfied in naturalistic environments. Later, researchers proposed echo cancellation strategies for estimating RIRs from reverberant speech signal second order statistics and using them to invert reverb effects. Many signal processing-based algorithms have been proposed with the goal of reducing processing latency versus the level of attenuation required. Later, many array processing solutions were addressed, such as spatio-temporal filtering, eigen decomposition, multi-channel system identification, which used multiple microphones to capture spatial information used to pick the direct path and reduce the energies in reflections from recorded speech signals. 

To address speech denoising tasks, time-frequency masking approaches were first introduced. Later, this technique was adapted to address speech dereverberation too. The ideal binary mask (IBM) has been shown in studies to improve speech intelligibility in reverberant speech. To compute the speech presence probability (SPP) in each time-frequency cell of a reverberant spectrogram, IBM treats the direct path and early reflections as the target speech and the rest as the masker. This mask is then applied to the reverberant spectrogram in order to resynthesize the dereveberated speech signal. IBM-based approaches were considered to have a hard decision boundary, which was the cause of musical noise artifacts in synthesized speech. To address these concerns, the hard decision boundaries were converted to soft boundaries to create an ideal ratio mask (IRM). Initially, IBM and IRM were applied to the magnitude responses of the distorted spectrograms, and the unaltered phase of the reverberant spectrogram was used to reconstruct the enhanced speech signal. This method has assisted many speech back-end systems in improving their performance. However, because the phase distortions compensated for the magnitude response improvements, this approach was unsuitable for improving the perceptual quality of the speech signal. To address this issue, researchers proposed the complex ideal ratio mask (CRM) \cite{crm1,crm2,crm3}, which estimates a complex time-frequency (TF) mask with the goal of enhance both the magnitude and phase of the reverberant speech complex spectorgram.

With advancements in deep neural networks (DNN) \cite{InceptionNet,ResNet,DenseNet,LSTM} most speech researchers have begun developing speech denoising and dereverberation strategies to estimate TF-masks. Many researchers have also looked into denoising autoencoders, recurrent networks, gated recurrent units (GRU) \cite{GRU}, convolution networks and their variants such as temporal convolution networks (TCN) \cite{tcn_sa,tcnn}, fully convolutional networks (FCN) \cite{fcn,unet}, gated convolution networks, and so on. Unlike most signal processing methods, deep neural networks can learn patterns for either denoising or dereverberation and generalize them to larger unseen scenarios with the help of nonlinear optimization. In recent years, many approaches in the field of image processing have been adapted to address speech enhancement for enhancing time-frequency spectorgrams. Self-attetnion networks (SAN) \cite{conv_san,pandey_san} and generative adversarial networks (GANs) \cite{gan,fsegan,cgan_se_sv} are two such deep learning techniques that have attracted attention. SANs are known for their performance in sequence-to-sequence tasks, such as machine translation. Self attention (SA) \cite{attn1,attn2} is a selective context aggregation, in which the predictions are computed based only on a subset of the input sequence it learns to attend. This ability to learn the regions of a sequence to attend for better estimations have been successfully used for many speech applications, including speech enhancement. On the other hand, GANs have been known for their ability to generate convincing unseen images when trained on natural images. Lately, researchers have proposed many improvements to the GAN architecture, which help improve the quality of the generated images by enhancing the fine details in the image. Many researchers are still exploring the potential of this approach for speech enhancement in the presence of additive noise. Some of the GAN approaches that showed promising improvements in the perceptual quality of speech are speech enhancement GAN (SEGAN) \cite{segan}, frequency-based speech enhancement GAN (FSEGAN) \cite{fsegan}, MetricGAN \cite{metricgan}, etc. Current studies focus on various GAN approaches for speech denoising, but few have looked into the GANs' abilities for speech dereverberation  \cite{dereverb_gan1, dereverb_gan2}. In this study, we keep our focus on building a speech dereverberation system using self-attention and GAN. We use the REVERB challenge corpus \cite{ReverbChallenge1,ReverbChallenge2}, which is widely used to benchmark improvements in the field of speech dereverberation to evaluate our proposed system.

Our main contributions towards the proposed network are summarized as follows:
\begin{itemize}
  \item We propose a novel fully complex-valued generative adversarial network, SkipConvGAN, that uses time-frequency masking to enhance the complex spectrogram. Unlike other GAN-based systems that handle magnitude response or time-domain speech signals using real-valued networks, our method incorporates complex-valued networks into the design of both the generator and discriminator of the proposed network.
  
  \item We replace the skip connections in the generator network with the proposed complex-valued convolutional modules, i.e., Skipconv blocks (SB) to bridge the semantic gap between feature maps exchanged between the encoder and decoder of the generator network and strengthen feature representation.
  
  \item We propose a complex-valued time-frequency self-attention (TF-SA) module that attends to features in both time and frequency dimensions while also preserving the interdependence of the real and imaginary components of the intermediate complex-valued feature maps.
  
  \item We train the proposed network with feature loss in addition to adversarial loss utilizing the complex-valued patch-discriminator network as a feature extractor.
  
\end{itemize}

The remainder of the paper is organized as follows: Section II summarizes the solution for speech dereverberation using a time-frequency complex ratio mask (CRM). Section III describes the components of our proposed SkipConvGAN network in detail. Section IV covers the training procedure for the proposed network. Section V describes the corpus and introduces signal processing and deep learning based algorithms against which our proposed system is evaluated. Section VI discusses the improvements in speech quality metrics achieved by various enhancement systems. In section VII, we summarize our findings and provide directions for further work.

\section{Signal Model and Problem Statement}
\label{sec:Model}
In this section, we first discuss the signal model for reverberant speech as well as the complex ratio masking (CRM) approach taken towards the speech dereverberation. For a given acoustic environment, speech captured by an omni-directional microphone is distorted by the reflections from the environment and the presence of any background noise which can be modeled as:
\vspace{-1em}
\begin{equation} 
\begin{aligned}[b]
y(t) &= s(t)*h(t) + n(t) = \sum_{t=0}^{L}{s(t)h(L-t)} + n(t)
\label{eq:reverb}
\end{aligned}
\end{equation}

\vspace{-0.5em}\noindent where `$\ast$' stands for the convolution operator, $s(t)$ is the clean speech from the source, $h(t)$ is the room impulse response (RIR) from the source to the microphone, $n(t)$ is the background noise, and $y(t)$ is the signal as observed by a distant microphone. The RIR can be divided into two components: (i) $h_d(t)$, which represents direct sound from the source, and (ii) $h_r(t)$, which represents all early and late reflections. Thus, the observed reverberant speech can be represented as follows:
\vspace{-0.5em}
\begin{equation} 
\begin{aligned}[b]
y(t) &= s(t)*h_d(t) + s(t)*h_r(t) + n(t) \\
y(t) &= x(t) + r(t) + n(t)
\label{eq:reverb}
\end{aligned}
\end{equation}

\noindent where $x(t)$ is the anechoic speech, and $r(t)$ is the reflections from the environment. The goal of this study is retrieve $x(t)$ from the observed speech signal $y(t)$. In practical scenarios, there always exists a time delay between the clean speech $s(t)$ and the anechoic speech at the microphone $x(t)$ which is directly proportional to the distance between source and the microphone. Due to the fact that we are dealing with a single channel, we can discard this delay by aligning the clean speech with the microphone signals.
The relation between observed reveerberant speech and its anechoic counterpart in Eq-(\ref{eq:reverb}) can be expressed in frequency domain as can be represented as:

\begin{equation} 
\begin{aligned}[b]
Y(t,f) &= X(t,f) + R(t,f) + N(t,f)
\label{eq:reverb_freq}
\end{aligned}
\end{equation}

\noindent where $Y(t,f)$, $X(t,f)$, $R(t,f)$, and $N(t,f)$ represent the complex spectrograms obtained by taking STFT of observed reverberant speech, anechoic speech from the source at the microphone, reflections, and background noise respectively. Thus, in frequency domain, we aim at retrieving $X(t,f)$ from $Y(t,f)$ by estimating a time-frequency (T-F) mask using the proposed network. This T-F mask is applied to the reverberant speech spectrogram to estimate the anechoic speech spectrogram as follows:  

\vspace{-0.5em}
\begin{equation} 
\begin{aligned}
\begin{split}
\hat{X}(t,f) &= \Phi_{model}(Y(t,f))Y(t,f) = M(t,f)Y(t,f),\\
&= (M_r\cdot Y_r-M_i\cdot Y_i) + j(M_r\cdot Y_i+M_i\cdot Y_r) 
\end{split}
\label{eq:masking}
\end{aligned}
\end{equation}
\vspace{-0.5em}

where, $M(t,f)$ is CRM estimated by the proposed network  ($\Phi_{model}$). $M_r$, $M_i$, $Y_r$, $Y_i$ are the real and imaginary components of estimated CRM, complex spectrogram of reverberant speech respectively. 
	
\section{Algorithm Description}
\label{sec:Algo}
The proposed network is based on the idea of recovering lost formant structure in captured reverberant speech signals using a generative model. For this purpose, we develop a GAN-based network, SkipConvGAN. This section begins by introducing the overall architecture of SkipConvGAN and then delves into the specifics of each module respectively.

\begin{figure*}[t!]
  \centering
  \includegraphics[width=\textwidth]{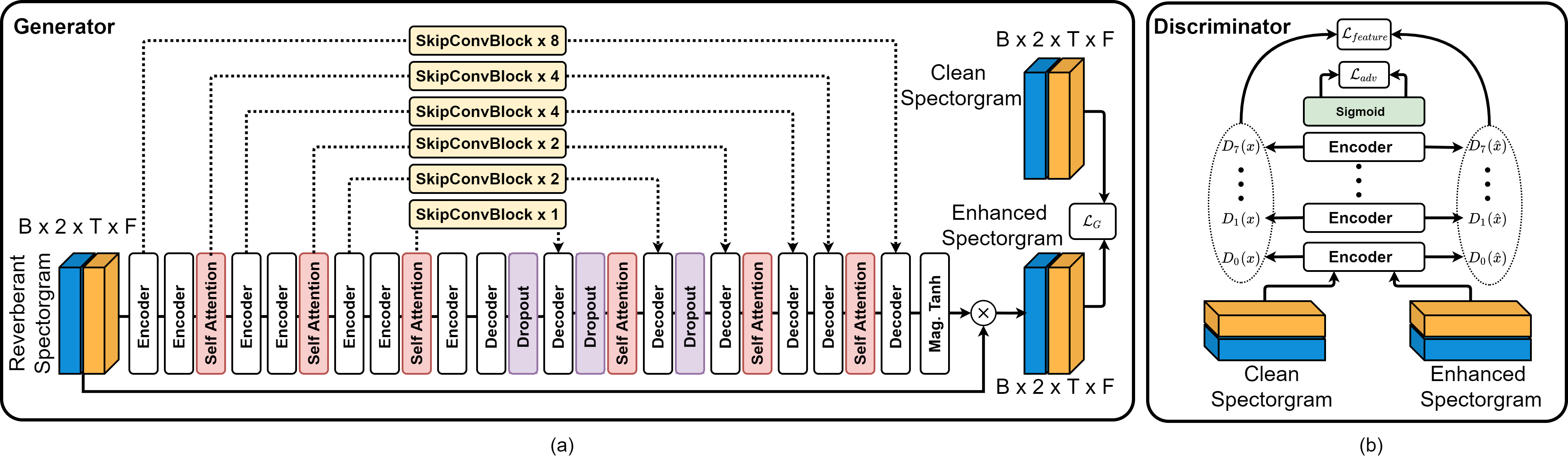}
  \caption{(a) Complex-valued generator network for the proposed network, (b) Complex-valued patch discriminator network for the proposed network.}
  \label{fig:Network}
\end{figure*}

\subsection{Overall Architecture}
\label{sec:Network}
The overall architecture of the proposed SkipConvGAN is depicted in Fig. \ref{fig:Network} which comprises of a generator network and a patch-discriminator network. The generator network is composed of three major components: (i) a fully convolutional complex-valued encoder-decoder network (Cplx-UNet) as a backbone, (ii) complex-valued SkipConvBlocks (SB) within the skip connections connecting the encoder and decoder blocks of the generator network, and (iii) complex-valued time-frequency self-attention (TF-SA) modules. The decoder network, on the other hand, is designed using a series of encoder blocks followed by a `sigmoid' activation at the end for binary classification. We train both generator and discriminator networks concurrently with contradictory losses. However, we discard the discriminator network during evaluation and use only the generator network as our monaural speech dereverberation system. The next subsections outline each component, followed by the proposed GAN's training approach.

\subsection{Complex-valued Encoder/Decoder Layer}
\label{sec:Conv}
A complex convolution serves as a building block of a complex-valued encoder/decoder layer used in the proposed network.  The idea of incorporating complex-valued convolutions into neural networks stems from the consistent improvement in magnitude responses of noisy and reverberant speech signals demonstrated by CNNs, as well as the fact that complex numbers can produce meaningful representations and outcomes when dealing with time-frequency complex spectrograms. In a conventional convolution layer, the convolution operation is computed by sliding over the kernel matrix over the input matrix and performing a point-wise multiplication. In a complex-valued generalization of a convolution layer, both the kernel and the input are complex-valued matrices. The convolution operation is unchanged. However, the distinction is due to the nature of multiplication operation for complex number in general. For example, the following explains the complex convolution operation over a complex input `$U$' given by $(U_r + jU_i)$, and complex kernel `$W$' given by $(W_r + jW_i)$, to produce an output `$Z$'.

\vspace{-0.5em}
\begin{equation} 
\begin{aligned}[b]
Z &=W\ast X = (W_r+jW_i)\ast(U_r+jU_i) \\ 
&= (W_r\ast U_r +W_i\ast U_i) + j(W_r\ast U_i-W_i\ast U_r)
\end{aligned}
\label{eq:cmplxconv}
\end{equation}

To implement this operation using real-valued convolutions, the real and imaginary components of the input and kernel matrices are stacked as additional channels and convolved in alternating orders to compute the output's corresponding real and imaginary components, see Fig. \ref{fig:complexconvolutions}(a). For instance, the following steps are performed sequentially to compute the real component of a complex-valued convolutional layer output: (i) The real part of the input is convolved with the real part of the kernel; (ii) The imaginary part of the input is convolved with the corresponding imaginary part of the kernel, and (iii) the outcomes of the above two operations are added. As indicated by the colored arrows in Fig. \ref{fig:complexconvolutions}(a), the imaginary component of the output is computed in the same manner. 
Each encoder layer in the proposed network is constructed using a `complex convolution', followed by `complex batch-normalization' and `complex activation', as illustrated in Fig. \ref{fig:complexconvolutions}(a). A decoder layer is also constructed in similar way. However, the `complex convolution' is substituted for `complex transpose convolution'. We highly encourage readers to refer \cite{cmplx1, cmplx2, cmplx3, pandey_cmplx} to gain a better understanding of how these complex-valued functions operate.

\subsection{Complex-valued SkipConv Blocks (SB)}
\label{sec:Skip}
As previously stated, the proposed network is built using a fully convolutional complex-valued encoder-decoder network (Cplx-UNet), which is a standard U-Net \cite{unet, image2image} framework with traditional convolution blocks substituted with complex equivalents. A U-Net is an encoder-decoder based fully convolutional network (FCN) that was originally developed for image segmentation task \cite{unet,fcn,image2image}. Recently, the speech community has seen a surge in the adoption of this architecture in conjunction with magnitude responses or complicated spectrograms for a variety of applications. A U-Net is composed of three critical components: an encoder, a decoder, and skip connections. The encoder of the network employs a series of convolutional layers to progressively extract higher dimensional information by processing local information from preceding layers and downsampling for subsequent layers. On the other hand, the network's decoder uses transpose convolution layers to upsample the features and generate higher-resolution images based on features learned from prior layers. To optimize image synthesis, higher-dimensional information acquired at each encoding layer is passed via `skip-connections' to the appropriate decoding layer and used to aid the decoder during image synthesis.

\begin{figure}[htb!]
\centering
  \includegraphics[width=0.8\linewidth]{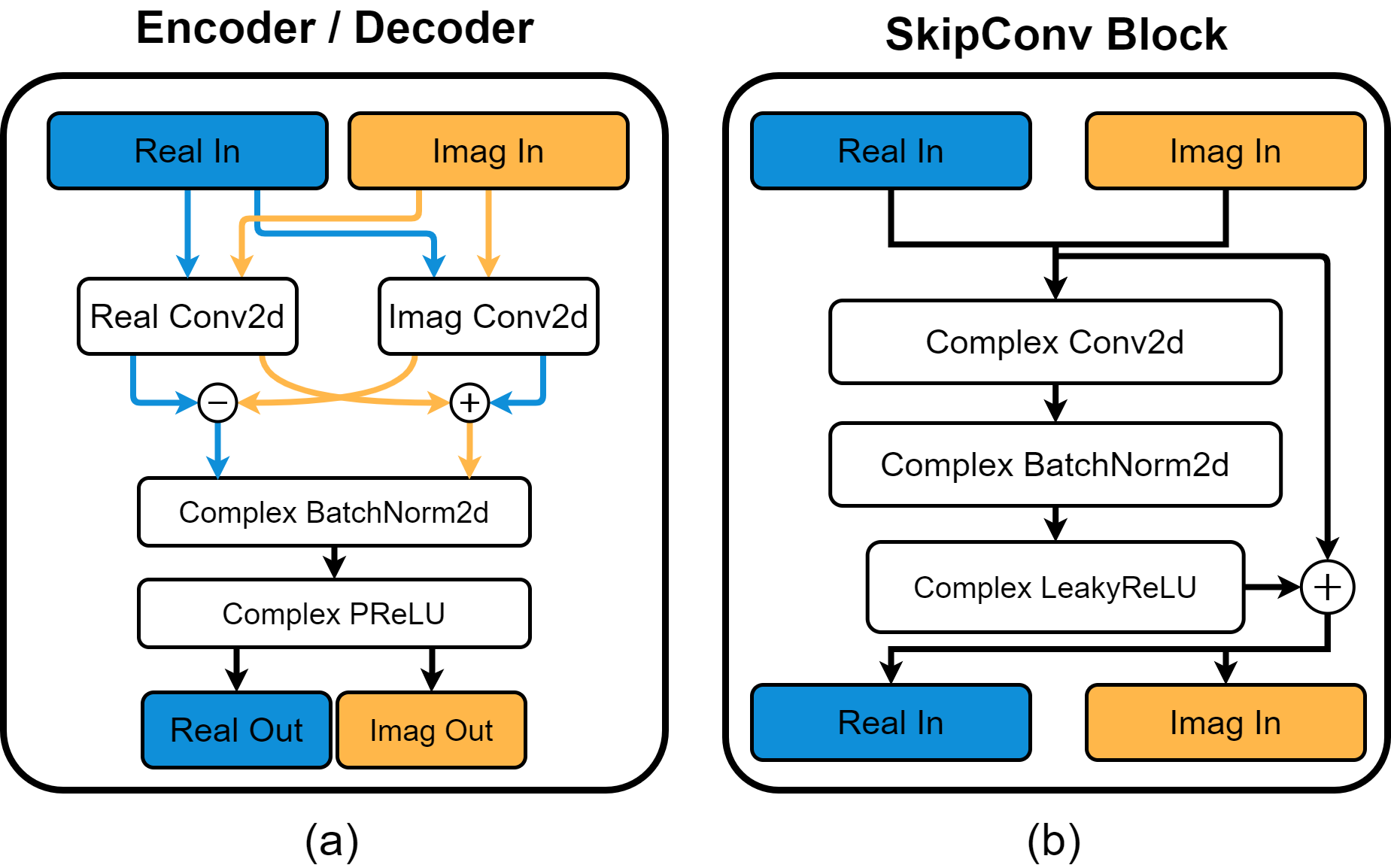}
  \caption{(a) Architecture of an Encoder/Decoder block, and (b) Architecture of an complex SkipConv block used in the proposed network.}
\label{fig:complexconvolutions}
\end{figure}

Although skip-connections have been shown to be efficient in building robust image-to-image systems, a recent study \cite{MultiResUNet} has identified a possible semantic gap in feature maps shared between the encoder and decoder using skip-connections. For instance, consider a spectrogram being fed into this network as an input. The output of the first encoder block is shared with the final decoder block, which combines the encoder's shared information with highly processed feature maps from the preceding decoder block to learn a mapping to a higher-resolution image. Combining these two incompatible set of features may hinder the decoder layer's ability in image reconstruction. To this end, we propose adding convolution layers within each skip link to compensate for these incompatibilities by transforming the shared feature maps between encoder and decoder layers to make them more intuitive for the corresponding decoder layer. Our earlier study \cite{skipconvnet} demonstrated that integrating SkipConv blocks (SB) within skip-connections significantly improves the learning ability of a given FCN network by reducing semantic gaps within the input features at each layer of the decoder. In the proposed network, we use a complex-valued version of SkipConv blocks (SB) where the real-valued components are replaced with its complex-valued equivalents. The architecture of a SkipConv block used in the proposed network is depicted in Fig. \ref{fig:complexconvolutions}(b). Each SB-block has a `complex convolution' followed with a `complex batch-normalization', a `complex activation' and a `residual connection'.

\subsection{Complex-valued time-frequency Self Attention (TF-SA)}
\label{sec:Attn}
\begin{figure}[htb!]
  \centering
  \includegraphics[width=\linewidth]{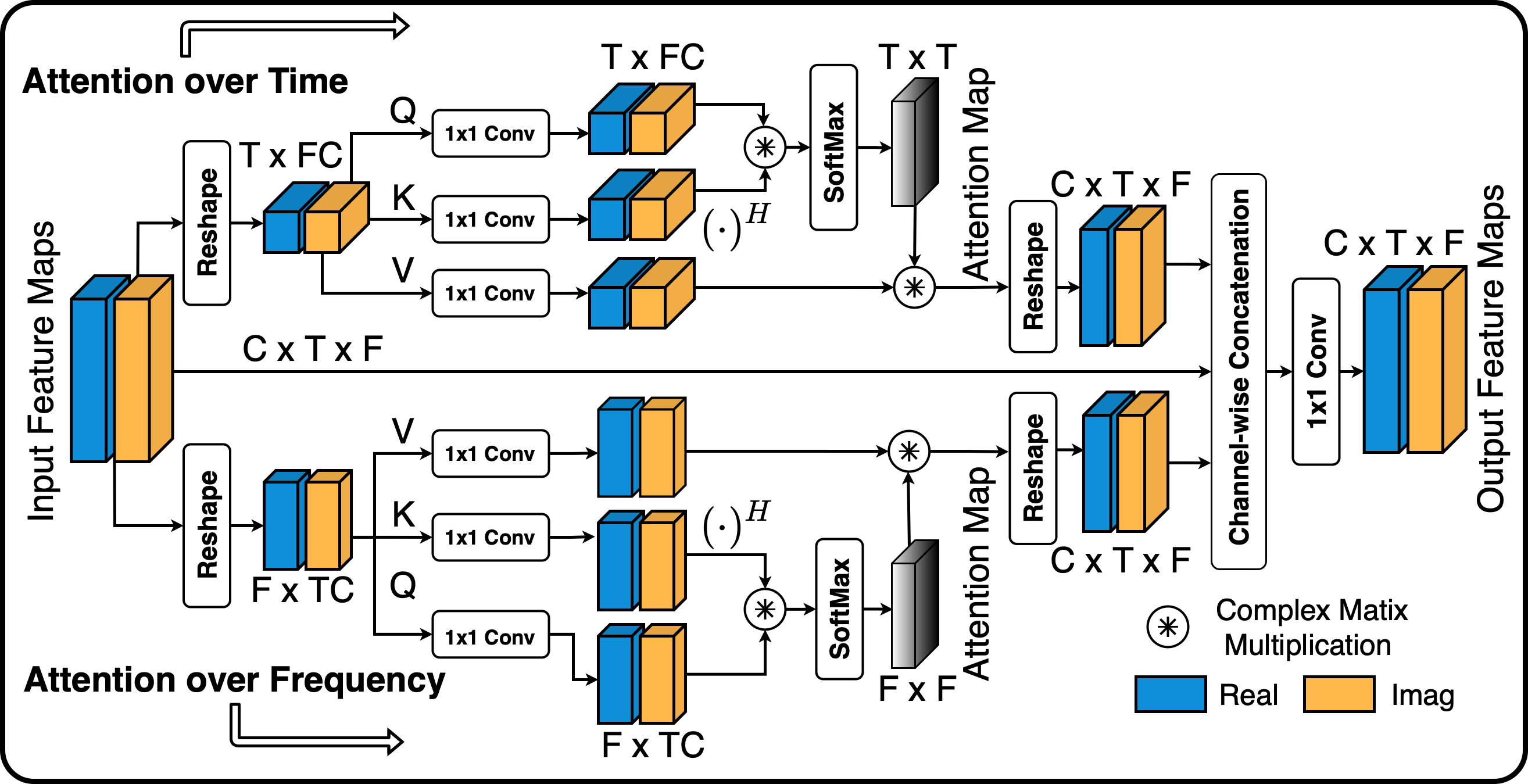}
  \caption{Self-Attention mechanism over time and frequency for complex domain.}
  \label{fig:cplxattention}
\end{figure}

Self-attention is a technique for relating different positions in a time sequence to compute a representation of the sequence that has been popular in recent years in a variety of speech applications. In speech applications, the majority of networks employ self-attention frameworks with real-valued networks that operate on magnitude responses or employ self-attention frameworks with feature-wise concatenation of real and imaginary components to handle complex domains. In this study, we extend the conventional self-attention mechanism to handle complex domain by proposing a complex-valued time-frequency self-attention (TF-SA) module that attends to features in both the time and frequency dimensions while maintaining the interdependence of the real and imaginary components of the intermediate complex-valued feature maps. The architecture of the proposed TF-SA mechanism is depicted in Fig. \ref{fig:cplxattention}. The TF-SA mechanism is composed of two branches, each of which computes attention in the time and frequency dimensions. We now briefly review the computation over the time dimension. Consider a set of complex-valued feature maps in a given layer of the network that consists of a sequence of 1-D frequency vectors with a total of $T$ frames, $U \in \mathbb{C}^{B{\times} C{\times} T{\times} F}$, where $\langle B \rangle$ denotes number of samples in a batch, $\langle C \rangle$ denotes the channels, $\langle T,F \rangle$ denotes the time and frequency dimensions respectively. We first reshape the feature maps to form a three dimensional tensor $U_t \in \mathbb{C}^{B{\times} T{\times} FC}$ to stack 1-D frequency vectors outputs from different kernel filters for real and imaginary parts of the feature maps. Next, we use 1-D complex convolution layer with a kernel size of $1{\times}1$ to project the sequence $U_t$ to different spaces to yield sequences query ($Q$), key ($K$) and value ($V$) as follows: 

\begin{equation} 
\begin{aligned}
(Q,K,V) = (W^QU_t,\:W^KU_t,\:W^VU_t) \;\; \in \mathbb{C}^{B{\times} T{\times} FC}\\
A = softmax(|QK^H|),\;\;\;\; A \in \mathbb{R}^{B{\times} T{\times} T}\;\;\;\;\;\;\;\; \\
SA_{time}(U) = AV = A(\Re{(V)}{+}j\Im{(V)})\;\;\;\;\;\;\;\;
\end{aligned}
\label{eq:attention}
\end{equation}

where $\Re{(\cdot)},\Im{(\cdot)}$ represent real and imaginary components of a complex number, and $W^{Q,K,V}$ indicates complex projection weight matrices for $Q$,$K$,and $V$ respectively. The attention map ($A$), is computed by complex dot-product operation followed by softmax on the magnitude as shown in Eq. \ref{eq:attention}. The resulting attention map takes a form of a 2-D matrix of dimensions `$T{\times} T$' for each sample in the batch where each row is a probability vector which represents the contribution of each frame to each other. For reverberant speech signals, we can anticipate that neighboring time frames will contribute the most, and that contributions will diminish as the distance between the time instances increases. At this point, the attention map is applied to the value sequence $V$ using complex matrix multiplication. Finally, this matrix is rearranged to generate output of attention over time, $SA_{time} \in \mathbb{C}^{B{\times} C {\times} T{\times} F}$. This is shown in the upper branch in Fig. \ref{fig:cplxattention}.

Likewise, we also compute attention over frequency  $SA_{freq}$ in the proposed TF-SA module since reverberation not only distorts the speech signal over time but also creates a smearing effect across frequency. This is shown in the lower branch of the Fig. \ref{fig:cplxattention}. The computation for attention over frequency remains the same as explained earlier for temporal attention with only a small change in the reshaping operation done at the first step of the process. Here we reshape the complex-valued feature maps ($U$) to form a three dimensional matrix $U_f \in \mathbb{C}^{B{\times}F{\times}TC}$ to stack 1-D time vectors from the outputs of different kernel filters for real and imaginary parts of the feature maps. 

\vspace{-0.2em}
\begin{equation} 
\begin{aligned}[b]
U_{c} = Concat(\{U, \;\;SA_{time}(U), \;\;SA_{freq}(U)\}) \\
\textrm{TF-SA}(U) = U_{c}W^O \;\;\;\;\;\;\;\;\;\;\;\;\;\;\;\;\;\;\;\;\;\;
\end{aligned}
\label{eq:TFSA}
\end{equation}

In conventional self-attention modules, the output of the attention layers are added to the input using a trainable scalar weight parameter. The network is allowed to learn the contributions of the attention layers as the training progresses. However, as shown in Eq. \ref{eq:TFSA} in the proposed TF-SA method, we concatenate the output of attention layers across time and frequency channels with the input ($U_{c}$) and employ a `$1{\times}1$' complex convolution layer ($W^O$) to allow the network to learn channel-specific weights, rather than combining the attention layers and input using a scalar that treats all channels equally.

\vspace{1em}
\subsection{Complex-valued Generator}
\label{sec:Gen}
As shown in Fig. \ref{fig:Network} (a), the generator network (Cplx-UNet+SB+TF-SA) is built on complex-valued FCN network (Cplx-UNet) using complex-valued skipconv blocks (SB) within the skip connections and complex-valued time-frequency self-attention mechanism (TF-SA) described in earlier sections. The generator network in the proposed network consists of seven encoder and decoder blocks as illustrated in Fig. \ref{fig:Network}(a). All complex-valued convolutions in the encoder and the decoder blocks use a kernel size of `$5{\times} 3$' with a stride of `$1{\times} 2$' in time and frequency dimensions respectively. The number of output channels for each encoder increases linearly as $\{1,16,32,64,128,256,512\}$. On the other hand, the number of output channels for each decoder decreases linearly in a similar fashion. However, the number of input channels for each decoder layer is doubled by the use of complex-valued skipconv blocks (SB) between the encoder and the decoder. Each encoder and decoder are connected using a series of complex-valued skipconv blocks (SB). The number of SB modules used to replace a skip connection is varied inversely with the depth of the encoder and the respective decoder layers within in the network i.e., $\{8,4,4,2,2,1\}$, see Fig. \ref{fig:Network}(a). This is based on the notion that deeper layers of the network dealing with high-level features require minimal change to close the semantic gap compared to the beginning layers. In addition, a total of six TF-SA modules are introduced at symmetrical locations between the encoder and the decoder layers of the proposed network, see Fig. \ref{fig:Network}. This cascaded connection of two encoder/decoder layers followed by a TF-SA module enables the network to gradually extract features and then learn to attend to key features in time and frequency at various stages of the network, resulting in the construction of a more accurate and efficient time-frequency (TF) mask. The generator takes in a complex spectrogram of reverberant speech signals, $Y(t,f) \in \mathbb{C}^{B{\times} 2{\times} T{\times} F}$ to estimate a complex time-frequency mask, $M(t,f) \in \mathbb{C}^{B{\times} 2{\times} T{\times} F}$. We then use the estimated TF-mask and input reverberant speech to compute the enhanced complex spectrogram, see Eq-\ref{eq:masking}.

\subsection{Complex-valued Patch-Discriminator}
\label{sec:Dis}
As shown in Fig. \ref{fig:Network} (b), the discriminator network (Cplx-Disc.) is built using a series of six complex-valued encoder layers. The first four encoder layers employ a kernel size of `$4{{\times}}4$' and a stride of `$2{\times}2$'. The final two encoder layers employ kernel sizes of `$3{\times}3$' and `$1{\times}1$' in the time and frequency dimensions, respectively. In addition, we also apply spectral normalization (SN) to the weights of the complex convolution within an encoder block as SN has been proved to achieve better performance by controlling weight ranges of the discriminator than the gradient penalty according to \cite{spectralnorm}. The discriminator network is designed to downsample the input complex spectrogram of size `$B{\times} 2{\times} T{\times} F$' to patches of size  `$B{\times} 2{\times} 16{\times} 16$' before proceeding to classify as real or fake. This is referred to as a patch-discrimination approach since it ensures that each of these smaller patches is trained to be categorized into the same class. This has been shown to improve the discriminator's robustness when training GANs.

\section{Training of proposed SkipConvGAN}
\begin{figure}[h!]
\centering
  \includegraphics[width=0.8\linewidth]{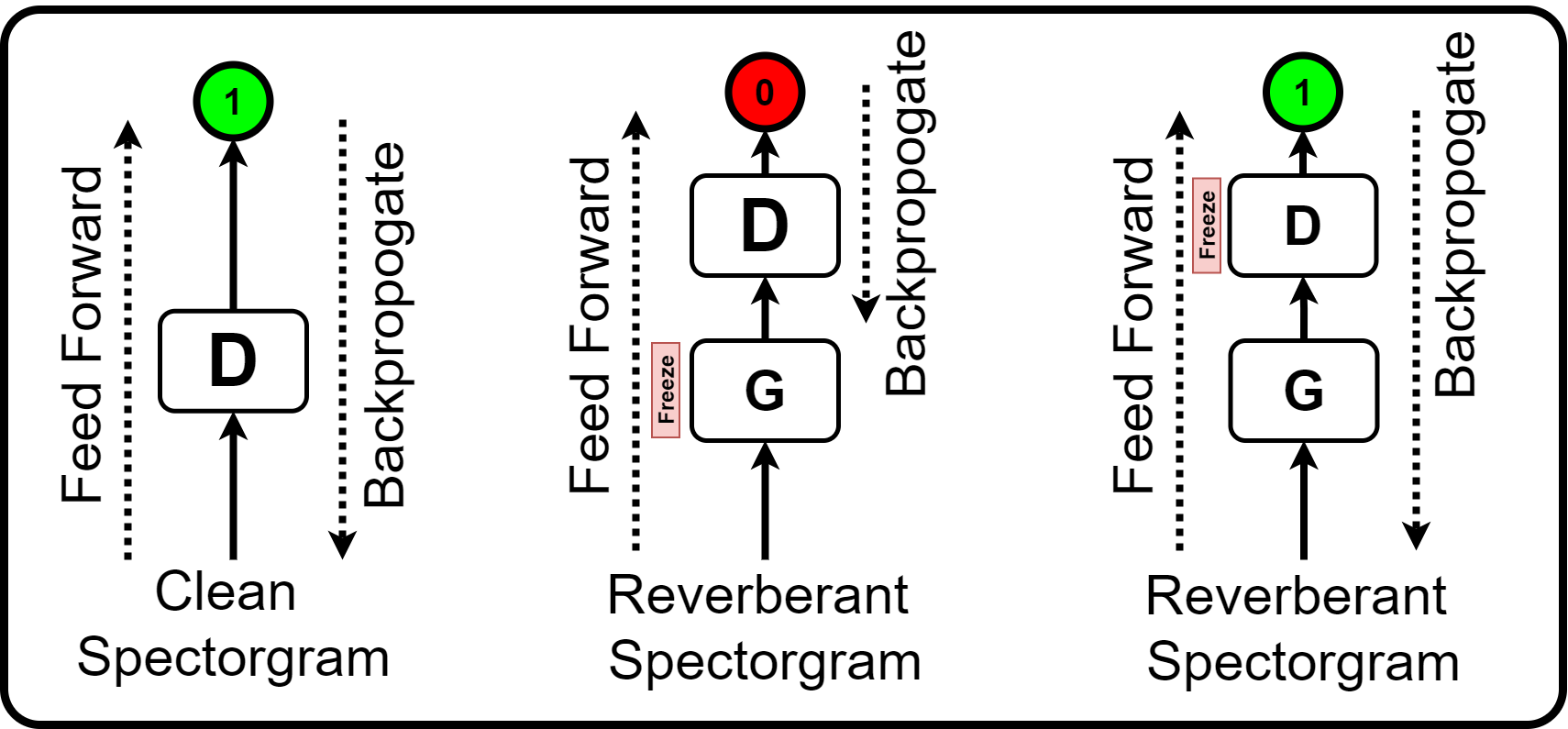}
  \caption{Training steps for generative adversarial networks}
\label{fig:GAN}
\end{figure}

This section delves into the details of adversarial training procedure of the proposed SkipConvGAN.  Similar to a traditional GAN approach for image generation, a generator network (Cplx-UNet+SB+TF-SA) learns to generate a complex time-frequency (TF) mask $M(t,f)$ to estimate an enhanced speech spectrogram $\hat{X}(t,f)$ from an input reverberant speech spectrograms $Y(t,f)$, see Eq. \ref{eq:masking}. The learning of the generator network is guided by an auxiliary discriminator network (Cplx-Disc.), which is simultaneously trained to distinguish between the enhanced speech spectrograms $\hat{X}(t,f)$ generated by the generator network from corresponding anechoic speech spectrogram $X(t,f)$ as fake and real respectively. This adversary learning task between competing generator and discriminator networks is achieved using mix-max optimization in conventional GANs. However, it has been discovered that this loss frequently causes the vanishing gradients problem during the learning process, making the GAN training process unstable \cite{DCGAN,cGAN,WGAN,InfoGAN}. To overcome this, we employ the least squares GAN (LSGAN) \cite{LSGAN} to train our proposed SkipConvGAN. This reduces the min-max optimization in a conventional GAN training to a simple minimization for both generator and discriminator networks as follows: 

\vspace{-1.5em}
\begin{equation} 
\begin{aligned}[b]
\min_{D} \mathcal{L}_{D} &= \frac{1}{2} \mathbb{E}[(D(X) - 1)^2] + \frac{1}{2} \mathbb{E}[(D(G(Y)) - 0)^2] \\ 
\min_{G} \mathcal{L}_{G} &= \frac{1}{2} \mathbb{E}[(D(G(Y)) - 1)^2]
\end{aligned}
\label{eq:gan_loss}
\end{equation}

where $\mathcal{L}_{G}$, $\mathcal{L}_{D}$ are the adversarial losses used to update the generator and the discriminator networks. The adversarial training process defers from the conventional training process followed for neural networks. The training process for GAN is carried out as a repetition of the following steps, see Fig. \ref{fig:GAN}. First, the discriminator is trained to distinguish between anechoic and enhanced speech spectrograms generated by the generator network using $\mathcal{L}_D$. While updating the discriminator, we fix the weights of the generator network. This ensures that the discriminator adapts its decision boundary across the estimated enhanced spectrograms over a large number of training samples in order to classify them as fake. After training the discriminator for a few iterations, the generator is trained using $\mathcal{L}_G$. The weights of the discriminator are fixed during generator training. This ensures that the decision boundary is fixed, allowing the generator to adapt by penalizing over all the training samples that are correctly classified by the discriminator. 

\vspace{-0.5em}

\subsection{Pre-training of the generator network}
The task of the generator network is far more difficult than that of the discriminator network in terms of learning an efficient complex-valued TF mask for a robust estimation of the enhanced speech spectrogram. If the generator network generates considerable artifacts, the discriminator's task becomes trivial which results in failure of GAN training. Therefore, it is critical to have the generator network start at a reasonable point in the beginning of the training, to ensure a proper convergence of the generator and discriminator’s loss functions. To this end, prior to starting the adversarial training, we first train the generator network with a combination of magnitude loss and complex-valued loss function as shown below:
\vspace{-1em}
\begin{equation} 
\begin{aligned}[b]
\mathcal{L}_{RI} = \frac{1}{TF}\sum_{t=0}^{T}\sum_{f=0}^{F}{\lVert\:
\hat{X}(t,f) - X(t,f) \:\rVert}_1 \;\;\;\\ 
\mathcal{L}_{Mag} = \frac{1}{TF}\sum_{t=0}^{T}\sum_{f=0}^{F}{\lVert\: |\hat{X}(t,f)| - |X(t,f)|\:\rVert}_1 \\
\mathcal{L}_{RI+Mag} = \lambda\mathcal{L}_{RI} + (1-\lambda) \mathcal{L}_{Mag} \;\;\;\;\;\;\;\;\;
\end{aligned}
\label{eq:featureloss}
\end{equation}
We train our generator network (Cplx-UNet+SB+TF-SA) for 20 epochs using the ADAM optimizer and a batch size of 16 and $\mathcal{L}_{RI+Mag}$ loss. Similar to \cite{loss}, the value of $\lambda$ is set to 0.3. The initial learning rate is set to 0.001 and is then divided by a factor of 10 when the loss doesn't reduce for two consecutive epochs.  

\subsection{Feature Loss}
\label{sec:Feat}
Alongside adversarial loss, we employ feature-based loss during the GAN training process inspired by \cite{medgan, denoising1, featloss_2, featloss_3}. Due to the fact that the layers in the discriminator network learn to extract features that distinguish anechoic and enhanced speech spectrograms, we consider the decoder network to be a trainable feature extractor and the intermediate features at different layers to be good representations of the spectrograms. Therefore, we use a feature loss by the discriminator to minimize the L1 distance between the extracted features for clean and enhanced spectrograms as shown in Eq. \ref{eq:featureloss}, to train our generator network. 
\vspace{-0.5em}
\begin{equation} 
\begin{aligned}[b]
\mathcal{L}_{feature} = \frac{1}{L}\sum_{i=0}^{L}{\lVert\: D_i(X) - D_i(\hat{X}) \:\rVert}_1
\end{aligned}
\label{eq:featureloss}
\end{equation}

where, `$L$' is the number of layers within the discriminator network, `$D_i(X),D_i(\hat{X}) $' are the intermediate feature representations of clean and enhanced spectrograms respectively. This avoids the generator network from over-training on the output statistics of a discriminator network while simultaneously encouraging it to estimate enhanced speech spectrograms with no artifacts. As a result, our formulation of feature loss promotes the convergence of GANs.

\subsection{GAN training}
We jointly train the pre-trained generator network and the discriminator network as an adversarial framework with a learning rate of $10^{-4}$ for both the generator and the discriminator networks. We use the following loss function for training the generator network in adversarial framework:

\begin{equation} 
\begin{aligned}[b]
\mathcal{L}_{Gen} = \alpha \mathcal{L}_{G} + \beta \mathcal{L}_{RI+Mag} + (1-\alpha-\beta) \mathcal{L}_{feature}
\end{aligned}
\label{eq:SkipConvGANloss}
\end{equation}

The weight decay was set to 0.001 for the discriminator and 0.0001 for the generator. Similar to the pre-training of the generator network, a batch-size of 16 is used to train SkipConvGAN with ADAM optimizer for 30 epochs. The value of $\alpha$ and $\beta$ are set to 0.4 and 0.3 respectively. For training samples, input reverberant time-domain signal, a 512-point short-time Fourier transform (STFT) with a window size of 32 ms and a window shift of 8 ms is applied to each frame, followed by the optimal smoothing pre-processing described in the later section. Subsequently, we split each utterance's complex speech spectrogram is divided into batches of 257 consecutive frames to create complex spectral images of size `$257{\times} 257$'.

\subsection{Optimal Smoothing Pre-Processing}
\label{sec:OSM}
Prior to training the proposed network, the training samples are pre-processed using the optimal smoothing pre-processing module based on minimum statistics for noise power spectral density (PSD) estimation \cite{minStats}. This approach has been shown to improve the performance of deep learning systems in general \cite{skipconvnet}. The approach \cite{minStats} was based on the observation that even during speech inactive regions, noise PSD estimate often decays to values which are representative of the noise power level. Thus, \cite{minStats} estimates the noise PSD by tracking the minimum power within a finite window large enough to bridge high power speech segments and averaging them using a time-frequency varying smoothing filter to avoid leakages of speech information. For reverberant speech signals, following the same concept would result in estimating the PSD of late reverberations. Although using \cite{minStats} approach is focused on estimating noise PSD (late reflections in our case), we only use the computed optimal smoothing parameter to average the late reflections from the training samples. Due to the fact that different samples in the training data set will contain varying amounts of late reflections, employing optimal smoothing to average out late reflections enables deep learning algorithms to model more effectively. 

\vspace{-0.5em}
\begin{equation}
    \centering
    \label{eq:opt_1}
    P(t,f) = \alpha(t,f)P(t{-}1,f) + (1-\alpha(t,f))|Y(t,f)|^2
\end{equation}

To compute the time-frequency varying optimal smoothing parameter $\alpha(t,f)$, we consider the first order smoothing equation for the PSD $P(t,f)$ from a given complex reverberant speech spectrogram $Y(t,f)$ as shown in Eq. \ref{eq:opt_1}. Similar to \cite{minStats}, we want to be as close as possible to the late reverberation and noise PSD $\sigma^2_{rn}(t,f)$ our objective is to minimize the conditional mean square error: 
\begin{equation}
    \centering
    \label{eq:opt_2}
    E\{(P(t,f)-\sigma^2_{rn}(t,f))^2 \lvert P(t{-}1,f)\}
\end{equation}

\begin{figure}[h]
    \centering
    \includegraphics[width=\linewidth, height=3.4cm]{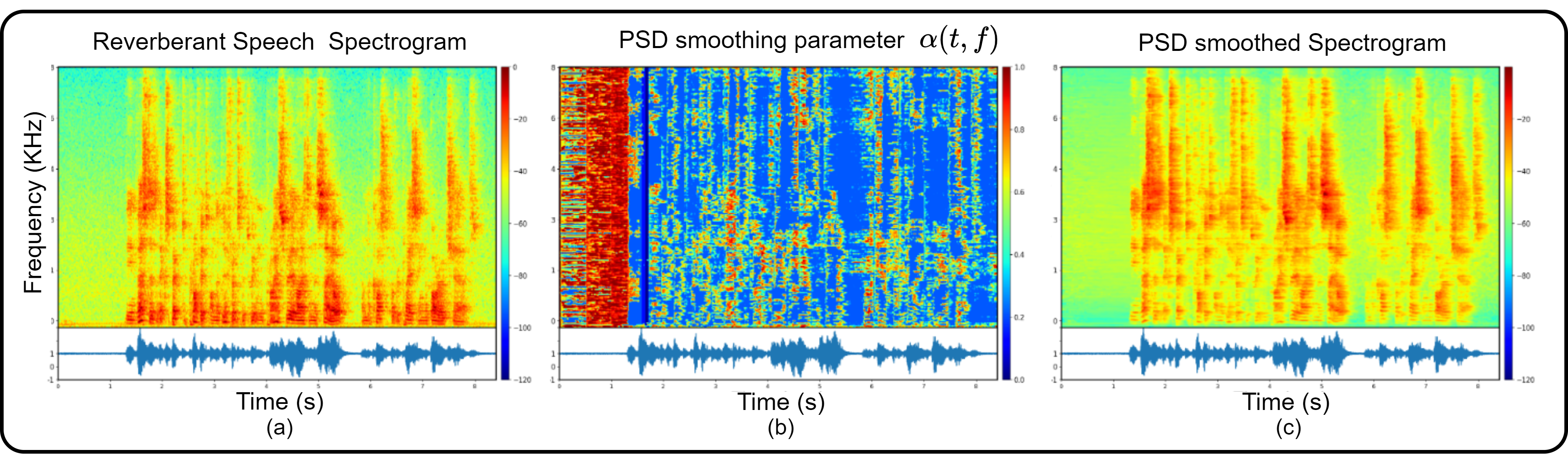}
    \caption{(a) Magnitude responses of a reverberant speech signal, (b) Time-Frequency varying PSD smoothing parameter, and (c) Magnitude response of pre-processed reverberant speech using optimal smoothing}
    \label{fig:Optimal_smoothing}
\end{figure}

During speech inactive regions, we assume psd of direct path $\sigma^2_x(t,f)$ to be zero and $E\{|Y(t,f)|^2\}$ represents late reflections and noise $\sigma^2_{rn}(t,f)$. Using the following assumptions and Eq. \ref{eq:opt_1} in Eq. \ref{eq:opt_2} and setting the derivative with respect to $\alpha(t,f)$ to zero yields the time-frequency varying optimal smoothing parameter as:
\begin{equation}
    \centering
    \label{eq:optimal_alpha}
    \alpha(t,f) = 1\big/(1+[P(t{-}1,f)/\sigma^2_{rn}(t,f){-}1]^2)
\end{equation}
For a detailed derivation of this optimal smoothing parameter, we recommend \cite{minStats}. In addition, the smooth PSD values below ‘-120 dB’ for all
training samples are clipped to have a constant dynamic range. Fig. \ref{fig:Optimal_smoothing} shows the original and smoothed versions of PSDs for a reverberant speech signal. 

The optimal smoothing parameter adapts itself accordingly for active and silent (late reflections in our case) regions of speech, see Fig. \ref{fig:Optimal_smoothing}(b). Thus, we propose to use this PSD smoothing strategy to pre-process the reverberant speech spectrograms before being fed to the proposed system for training.

\section{Experimental Setup}
\label{sec:Exp}

\subsection{The REVERB Challenge Corpus}

The Reverb Challenge corpus \cite{ReverbChallenge1, ReverbChallenge2} consists of simulated and real recordings of a stationary speaker captured with single-channel (1-ch), two-channel (2-ch), and eight-channel (8-ch) microphones sampled at 16 kHz. The corpus is divided into train, development, and evaluation sets, each of which contains non-overlapping speakers and speech utterances. The corpus's training set is composed of 7861 clean speech utterances from the WSJ0 corpus, with an average duration of 7.5 seconds. To simulate reverberant speech utterances with a signal-to-noise ratio of 20 dB, a total of 24 measured room impulse responses and pre-recorded background noise are used. The impulse responses used in the simulations have reverberation times ranging from 0.2 to 0.8 seconds. To train our systems, we only use simulated speech utterances provided by the corpus as training data, without augmenting it with any simulated data. The corpus's development and evaluation sets are used to fine-tune the systems and evaluate their performance, respectively. Unlike the training set, the corpus's development and evaluation sets comprises of a mix of simulated data (\textit{SimData}) and real recordings (\textit{RealData}) to ensure the systems are tested for: (i) robustness in realistic conditions, and (ii) robustness against a wide range of reverberation conditions. The development set's \textit{SimData} consists of 1484 speech utterances from the WSJCAM0 corpus \cite{wsjcam0} that are convolved with RIRs measured in three different rooms of different volumes (small, medium, and large) at two different speaker to microphone array distances (near, far) to generate reverberant speech utterances. The recorded background noise is stationary diffuse noise mainly caused by the air-conditioning systems in the rooms. Thus, the recordings tend to have relatively large low-frequency energies. Similarly, the \textit{SimData} for the evaluation set is generated under the same conditions and contains 2176 speech utterances.

The \textit{RealData} of the development and evaluation set includes 179 and 372 speech utterances from the MC-WSJ-AV corpus \cite{mcwsjav}, which incorporates utterances spoken by human speakers in a single noisy and reverberant room, respectively. The room used to record the \textit{RealData} differs from the room used to simulate the \textit{SimData}. Furthermore, for near and far conditions, the distance between the human speaker and the microphone array is varied to 100 cm and 250 cm, which also differs from the distances in the \textit{SimData}. This ensures that systems are evaluated for robustness against mismatching environmental conditions. The following section describes the architectural design and training details of the systems used to compare the proposed network.

\subsection{Comparison Systems}
\label{sec:Baselines}
We compare our proposed SkipConvGAN network to existing signal processing-based approaches to speech dereverberation, such as WPE and LPC residual enhancement, as well as deep learning-based approaches to speech enhancement, such as SkipConvNet, deep complex convolutional recurrent network (DCRNN), and GAN-based approaches, such as FSEGAN and MetricGAN. The next subsections discuss the architecture and optimization techniques used to design and train the systems.

\subsubsection{WPE speech dereverberation}
For comparing the proposed architecture, we use the weighted prediction error (WPE) dereverberation algorithm as a primary baseline. WPE \cite{WPE1,WPE2} uses long-term linear prediction to estimate the impact of RIRs from reverberant speech signals. This algorithm has been traditionally used to improve the performance of distant-talk automatic speech recognition (ASR) systems, by improving speech quality. This application is the motivation behind evaluating the speech enhancement capability of the WPE algorithm. The algorithm assumes that the early reflections of a RIR, in addition to the direct part, are both necessary and sufficient for a higher recognition rate. Furthermore, late reflections are considered to be diffused and are the cause of poor speech intelligibility. We use a three-sample delay and a ten-tap filter size configuration for WPE and run a total of 15 iterations on each speech utterance to estimate the filter weights that will be used to dereverberate the speech signal. Additionally, WPE is designed to be compatible with multi-channel data in order to improve filter parameter estimation by using spatial information from several channels. In this study, we compare the performance of single-channel and multi-channel WPE with MVDR beamforming and MMSE postfiltering with the proposed network

\subsubsection{LPC residual enhancement}
This is an unsupervised approach for speech dereverberation that employs a two-stage model \cite{twostage_lpc} to address two types of reverberation-caused degradation: (i) coloration, which is a change in the relative amplitude of different frequencies in captured reverberated speech compared to its anechoic counterpart, and (ii) late reverberation, which is the effects of reflections that take more than 50 ms to reach the microphone compared to the direct path. In the first stage, an inverse filter is estimated from the reverberant speech signal to reduce coloration effects which help improve the signal-to-reverberation ratio. This is accomplished by recursively learning the inverse filter by maximizing the higher-order statistics (kurtosis) of LP residual signal of the reverberant speech signal. Later, in the second stage, a spectral subtraction approach is used to minimize the influence of late reverberation. In our study, we use a 256-tap inverse filter learned over 150 iterations for each reverberant speech utterance. The short-term fourier transform (STFT) of inverse-filtered speech is obtained by using hamming windowed frames of length 16 ms with 8-ms overlap for spectral subtraction in the second stage. Spectral subtraction is employed to estimate the power spectrum of clean speech signal with a maximum allowed attenuation of 30 dB to avoid artifacts while speech reconstruction.

\subsubsection{SkipConvNet}
This system uses deep learning approach to learn a generalized non-linear mapping from a reverberant speech log-power spectrum (LPS) to a corresponding anechoic speech LPS \cite{skipconvnet}. We compute STFT with a frame length of 512 samples and an overlap of 384 samples. We then compute LPS of a speech signal from an optimally smoothed LPS of the speech using optimal smoothing as mentioned in the earlier section. We only consider the lower half, since the STFT is symmetric. Later, the LPS of each utterance is divided into batches with 256 consecutive frames to form spectral images of size `$256{\times}256$'. We employ the architecture mentioned in \cite{skipconvnet} with seven layers of encoder-decoder pairs and skipconv blocks within the skip connections. All convolutions in the encoder and skipconv blocks use a kernel size of 5x5 with a stride of 2. Likewise, all convolutions in the decoder use transposed convolution with a kernel size and stride of 2. Unlike in \cite{skipconvnet}, number of filters used in the encoder and the decoder are reduced to make the network lighter and faster to train. We train the network to minimize the mean square error (MSE) between the estimated and corresponding clean LPS using Adam optimizer. A batch-size of 8 is used to train the network for 20 epochs. Finally, estimated LPS from the network is combined with the unaltered noisy phase to reconstruct the enhanced speech.

\subsubsection{DCCRN}
Deep complex convolutional recurrent network (DCCRN) is another deep learning-based approach that has gained popularity due to the improved speech quality demonstrated on a recent Deep Noise Suppression challenge (DNS)\cite{dccrn}. Unlike `\textit{SkipconvNet}', this system handles complex speech spectrograms using complex-valued convolutions and complex-valued LSTM layers. DCCRN, like the proposed study, aims to estimate a CRM and is optimized through signal approximation. The network's encoders and decoders have the same structure as described in previous sections. A complex-valued LSTM layer is implemented with two independent LSTM weights initialized for real and imaginary. The equation below is then used to compute the real and imaginary components of the LSTM output.

\vspace{-0.5em}
\begin{equation} 
\begin{aligned}[b]
Z_{rr} = LSTM_r(X_r); \quad \quad Z_{ir} = LSTM_r(X_i); \\
Z_{ri} = LSTM_i(X_r); \quad \quad Z_{ii} = LSTM_i(X_i); \\
Z_{out} = (Z_{rr} - Z_{ii}) + j(Z_{ri} + Z_{ir}) \quad
\end{aligned}
\label{eq:cmplxconv}
\end{equation}

where $LSTM_{r/i}$ are the weights of real and imaginary layers of the complex-valued LSTM. $X_{r/i}$ \& $Z_{r/i}$ are the real and imaginary channels of the input and output feature maps respectively. In this study, we employ the 'DCCRN-E' configuration provided in \cite{dccrn} since the masking strategy used in this configuration matches the configuration used in the proposed study. We train 'DCCRN-E' network to minimize Si-SDR loss using Adam optimizer to reduce the mean square error (MSE) between the estimated and clean spectrograms. The network is trained for 20 epochs with a batch size of 16.

\subsubsection{FSEGAN}
This method employs GAN to construct a robust automated speech recognition (ASR) system \cite{fsegan}. Most ASR systems train acoustic models by extracting mel-frequency cepstral coefficients (MFCC) as features from the magnitude response of speech signals. Thus, frequency-domain SEGAN (FSEGAN) was designed to improve only the magnitude response of the speech signal while ignoring its phase. FSEGAN, similar to the proposed network uses fully convolutional networks (FCN) to build the generator and the discriminator networks. The generator network consists of seven encoder and decoder pairs with a kernel filters of size `$4{\times} 4$' and a stride of 2. All convolution layers are followed by batch normalization and leakyReLU activation. however, batch normalization and activation are excluded for final decoder layer of the generator. To make FSEGAN a deterministic model, the intermediate layer output from the last encoder layer is extracted and used as a latent code, similar to SEGAN. The discriminator network, on the other hand, has four layers of convolutions with the same kernel filter and stride as the generator. The last layer of the discriminator, on the other hand, has kernel filters of size `$1{\times}8$' and sigmoid activation. The speech files from the training set of the corpus is chunked into audio of length 1.28 sec with a 50\% overlap. Next, a 512-point STFT is performed on these speech files to get to magnitude responses of shape `$128{\times} 128$'. We train this GAN-based approach for 80 epochs using the adversarial loss described in \cite{fsegan} with a batch size of 64.

\begin{figure*}[t!]
  \centering
  \includegraphics[width=\linewidth,height=7cm]{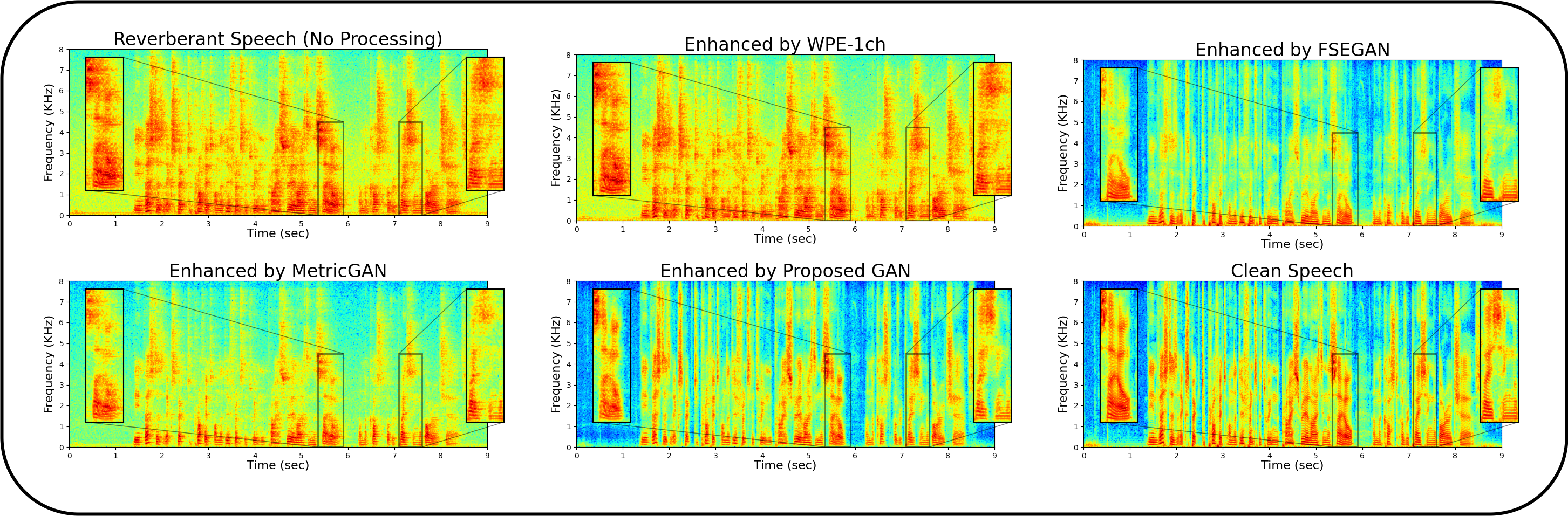}
  \caption{Magnitude responses of enhanced speech signals from various signal processing and deep learning based dereverberation systems}
  \label{fig:results}
\end{figure*}

\subsubsection{MetricGAN}
This GAN-based method optimizes the generator network in terms of one or more evaluation metrics. This is accomplished by associating the discriminator network to the desired evaluation metrics. In \cite{metricgan}, the networks were optimized for two distinct metrics: PESQ and STOI. However, our study focuses exclusively on optimizing the PESQ. The generator network of MetricGAN consists of two bidirectional-LSTM (BLSTM) layers of 200 nodes each, followed by two fully connected layers and LeakyReLU activations. To avoid musical noise artifacts during reconstruction, a sigmoid layer with a threshold of 0.05 is applied on the last layer to construct a time-frequency (TF) mask. Likewise, the discriminator network of MetricGAN is designed as a four-layer CNN with increasing kernel size. Following the convolution layers are three fully connected layers, LeakyReLU activation, and a single-node linear layer. Furthermore, the discriminator network is confined to 1-Lipschitz continuity by applying spectral normalization to all convolutional layers in the network. We train this GAN-based approach for 80 epochs using the adversarial loss described in \cite{metricgan} with a batch size of 128.

\begin{table*}[h!]
\centering
\caption{Ablation study of the proposed generator network's performance at different environments. Table shows improvements achieved in Average PESQ and SNR$_{fw}$.}
\scalebox{0.7}{
\renewcommand{\arraystretch}{1.9}
\begin{tabular}{lccccccc} 
\toprule
~ & \Large{\textbf{Loss func.}} & \multicolumn{3}{c}{\Large{\textbf{PESQ}}} & \multicolumn{3}{c}{\Large{\textbf{SNR}$_{fw}$ (dB)}}  \\ \hline
\Large{\textbf{room size}} & ~ & \Large{small} & \Large{medium}  & \Large{large} & \Large{small} & \Large{medium}  & \Large{large}\\ \midrule

\Large{\textbf{No Processing}} & \Large{-} & \Large{1.910} & \Large{1.314} & \Large{1.285} & \Large{7.401} & \Large{2.193} &\Large{1.255} \\
    
\Large{\textbf{Cplx-UNet}} & \Large{$\mathcal{L}_{RI+Mag}$} &\Large{2.085} & \Large{2.496} & \Large{2.127} & \Large{10.370} & \Large{8.897} & \Large{8.659} \\

\Large{\textbf{Cplx-UNet+SB}} & \Large{$\mathcal{L}_{RI+Mag}$} &\Large{2.965} & \Large{2.554} & \Large{2.304} & \Large{10.177} & \Large{9.090} & \Large{8.823} \\

\Large{\textbf{Cplx-UNet+SB+TF-SA}} & \Large{$\mathcal{L}_{RI+Mag}$} &\Large{3.126} & \Large{2.659} & \Large{2.498} & \Large{11.615} & \Large{10.312} & \Large{10.068} \\

\Large{\textbf{Cplx-UNet+SB+TF-SA+Cplx-Disc.}} & \Large{($\mathcal{L}_{G}+\mathcal{L}_{RI+Mag},\mathcal{L}_{D}$)} &\Large{3.214} & \Large{2.712} & \Large{2.654} & \Large{12.489} & \Large{11.269} & \Large{11.327} \\ \hline

\Large{\textbf{Proposed GAN}} & \Large{($\mathcal{L}_{Gen},\mathcal{L}_{D}$)} &\textbf{\Large{3.233}} & \textbf{\Large{2.766}} & \textbf{\Large{2.735}} & \textbf{\Large{12.807}} & \textbf{\Large{11.303}} & \textbf{\Large{11.573}} \\
\bottomrule
\end{tabular}}
\label{tab:AblataionStudy_RT60}
\end{table*}

\begin{table*}[b!]
\centering
\caption{Ablation study of the proposed generator network's performance at different microphone distances. Table shows improvements achieved in Average PESQ and SNR$_{fw}$.}
\scalebox{0.7}{
\renewcommand{\arraystretch}{2}
\begin{tabular}{lcccc} 
\toprule
~ & \multicolumn{2}{c}{\Large{\textbf{PESQ}}} & \multicolumn{2}{c}{\Large{\textbf{SNR}$_{fw}$ (dB)}}  \\ \hline
\Large{\textbf{mic distance}} & \Large{Far} (d=2m) & \Large{Near} (d=0.5m) & \Large{Far} (d=2m) & \Large{Near} (d=0.5m)\\ \midrule
    \Large{\textbf{No Processing}} & \Large{1.339} & \Large{1.667} & \Large{2.650} & \Large{4.582} \\
    \Large{\textbf{Cplx-UNet}} & \Large{2.529} & \Large{2.626} & \Large{8.671} & \Large{9.946} \\
    \Large{\textbf{Cplx-UNet+SB}} & \Large{2.513} & \Large{2.702} & \Large{8.683} & \Large{10.043}\\
    \Large{\textbf{Cplx-UNet+SB+TF-SA}} & \Large{2.579} & \Large{2.944} & \Large{9.628} & \Large{11.702}\\
    \Large{\textbf{Cplx-UNet+SB+TF-SA+Cplx-Disc.}} & \Large{2.714} & \Large{2.983} & \Large{10.568} & \Large{12.192}\\ \hline
    \Large{\textbf{Proposed GAN}} & \textbf{\Large{2.748}} & \textbf{\Large{3.073}} & \textbf{\Large{10.879}} & \textbf{\Large{12.909}}\\
\bottomrule
\end{tabular}}
\label{tab:AblataionStudy_Dist}
\end{table*}

\section{Results \& Discussion}
\label{sec:Results}
\subsection{Evaluation Metrics}
We evaluate the performance of the proposed and comparison systems using the speech quality metrics provided as part of the REVERB challenge corpus, which include: (i) cepstral distance (CD), (ii) log-likelihood ratio (LLR), (iii) frequency-weighted segmental SNR (FWSegSNR), (iv) perceptual quality (PESQ), and (v) signal-to-reverberation modulation energy ratio (SRMR). With the exception of SRMR, the computation of these metrics necessarily requires the use of a corresponding clean speech to compare the reverberant/enhanced speech for scoring purposes. As a result, improvements in all metrics for \textit{SimData} and only improvements in SRMR for \textit{RealData} on the evaluation set of the corpus are examined.

\begin{table*}[b!]
\centering
\caption{Improvements in speech quality achieved from different algorithms on SIMDATA and REALDATA OF The REVERB CHALLENGE EVALUATION SET.}
\scalebox{0.8}{
\renewcommand{\arraystretch}{2.0}
\begin{tabular}{p{7cm}cccccc} 
\toprule
~ & \multicolumn{5}{c}{\Large{SimData}} & \Large{RealData}  \\ \hline
~ & \Large{\textbf{CD}} & \Large{\textbf{SRMR}} & \Large{\textbf{LLR}}  & \Large{\textbf{SNR}$_{fw}$} & \Large{\textbf{PESQ}} & \Large{\textbf{SRMR}}\\ \midrule
    \Large{\textbf{No Processing}} & \Large{3.975} & \Large{3.687} & \Large{0.574} & \Large{3.617} & \Large{1.503} &\Large{3.180} \\
    \Large{\textbf{WPE}} (1-ch) & \Large{3.748} & \Large{4.220} & \Large{0.514} & \Large{4.864} & \Large{1.722} &\Large{3.978} \\
    \Large{\textbf{WPE}} (2-ch) & \Large{3.660} & \Large{4.500} & \Large{0.470} & \Large{5.350} & \Large{1.820} &\Large{4.480} \\
    \Large{\textbf{WPE+MVDR+MMSE}} (8-ch) & \Large{2.250} & \Large{5.390} & \Large{0.430} & \Large{10.310} & \Large{2.820} & \textbf{\Large{7.340}} \\
    \Large{\textbf{LPC residual enhancement}} & \Large{5.025} & \Large{5.827} & \Large{1.236} & \Large{5.210} & \Large{2.215} & \Large{5.953} \\
    \Large{\textbf{SkipConvNet}} & \Large{2.328} & \Large{4.852} & \Large{0.261} & \Large{10.746} & \Large{2.154} & \Large{7.060} \\
    \Large{\textbf{DCCRN}} & \Large{2.669} & \Large{4.628} & \Large{0.349} & \Large{9.369} & \Large{2.578} & \Large{5.075} \\
    \Large{\textbf{FSEGAN}} & \Large{2.628} & \Large{4.736} & \Large{0.319} & \Large{10.618} & \Large{2.729} & \Large{5.783} \\
    \Large{\textbf{MetricGAN}} & \Large{3.851} & \textbf{\Large{7.390}} & \Large{0.677} & \Large{3.908} & \Large{2.570} & \Large{7.205} \\ \hline
    \Large{\textbf{Proposed GAN}} & \textbf{\Large{2.318}} & \Large{5.887} & \textbf{\Large{0.234}} & \textbf{\Large{11.896}} & \textbf{\Large{2.911}} & \Large{6.355} \\
\bottomrule
\end{tabular}}
\label{tab:Results}
\end{table*}

\vspace{-0.5em}
\subsection{Ablation Study}
In this section, we present the findings of an ablation study conducted to determine the contributions of key components of the generator and discriminator network in the proposed system. All the systems in this ablation study are trained for 30 epochs using the same optimizer parameters. We use $\mathcal{L}_{RI+Mag}$ loss to train only the generator network and $\mathcal{L}_{G}$, $\mathcal{L}_{Gen}$ and $\mathcal{L}_{D}$ loss to train the GAN-based networks. The loss functions used to train each system in the ablation study are listed in Table \ref{tab:AblataionStudy_RT60}. The improvements obtained by combining each of the proposed components are demonstrated by comparing objective scores on the evaluation set of the REVERB CHALLENGE corpus.

\subsubsection{Generator Network}
We begin by constructing the proposed system using a simple complex-valued fully convolutional network (Complex-UNet) with six encoder and decoder pairs coupled using skip connections. We then subsequently incorporate additional modules. First, we add complex-valued SkipConv blocks (Complex-UNet+SB) into the skip connections which is intended to reduce the semantic gap between features from the encoder and the decoder, as described in our previous work \cite{skipconvnet}. Next, the self-attention module (Complex-UNet+SB+SA) proposed in this study (Section-\ref{sec:Attn}) is introduced to construct the generator network for the proposed GAN-based dereverberation system. The generator network is then trained in conjunction with the proposed complex-valued patch-discriminator network using adversarial loss (Complex-UNet+SB+SA+Cplx-Disc.). Finally, we train the generator network using adversarial and feature loss (Proposed GAN). 

 We evaluate the performance of these three systems for three different room dimensions in the evaluation set of the corpus. Table-\ref{tab:AblataionStudy_RT60}, shows improvements in average PESQ and frequency-weighted segmental SNR. Across different rooms, we observe that the base network with no additional modules (Cplx-UNet) achieves 48.76\% relative improvement in PESQ and 5.692 dB absolute improvement in SNR respectively when compared to the unprocessed reverberant speech. This suggests and ensures that robustness of the base network against unseen reverberant conditions. First, we integrate complex-valued skipconv blocks described in Section-\ref{sec:Skip} within the skip connections of the base network (Cplx-UNet+SB). From our previous work \cite{skipconvnet}, we infer that adding convolutional modules within the skip connections consistently improved the performance of the real-valued network trained on the magnitude spectrogram. However, with a complex-valued network, we observe a significant 16.62\% relative gain in PESQ, despite the absence of significant improvements in SNR. This indicates that reducing the semantic gap between the shared feature maps between the encoder and decoder components only benefits but does not impair the network's performance. Furthermore, the addition of the proposed complex-valued self-attention module to the network (Cplx-UNet+SB+SA) consistently improved the speech quality by an additional 5.88\% relative gain in average PESQ and 3.90 dB absolute improvement in average SNR respectively. In summary, we
 conclude that speech quality is progressively improved by adding the proposed components to the base network. Therefore, we use Cplx-UNet+SB+SA as the generator network in the proposed SkipConvGAN. 
 
 \subsubsection{Discriminator Network}
 In this section, we present the contributions of the discriminator towards the improvement in overall speech quality. We build the proposed SkipConvGAN using generator network (Cplx-UNet+SB+SA) and complex-valued patch-discriminator (Cplx-Disc.) described in Section-\ref{sec:Dis}.  We train the network using: (i) conventional adversarial loss (Cplx-UNet+SB+SA+Cplx-Disc.), and (ii) adversarial and feature loss (Proposed GAN). From Table-\ref{tab:AblataionStudy_RT60}, we observe that the generative training process achieves an additional 3.58\% \& 9.65\% relative improvements in average PESQ and SNR respectively. Likewise, from Table-\ref{tab:AblataionStudy_Dist}, we observe an additional 3.15\% and 6.70\% improvements in average PESQ and SNR across different microphone distances. This supports our intuition that generative training intended to restore the loss formant structure helps improve the overall speech quality. Furthermore, the with additional feature loss introduced to better guide the discriminator network in GAN training (Proposed GAN) further improves the quality across different reverberant conditions and microphone distances. Thereby, we suggest the use of the patch-discriminator along with feature loss for optimal performance. 

\subsection{Comparison with Baselines}
In this section, we extensively evaluate our proposed system using both `SimData' and `RealData' from evaluation set of the REVERB challenge corpus. Fig. \ref{fig:results} depicts spectrograms of enhanced speech signals obtained by various dereverberation systems mentioned in the study for a speech sample captured by a `far microphone' in `Room 2'. Similarly, Table \ref{tab:Results} lists the improvements in speech quality metrics achieved by all systems in the study.

We begin by revisiting the pre-processing module, which is used for preparing training samples for the proposed system. The pre-processing module is a two-step process that includes pre-emphasis and optimal smoothing of reverberant PSD. When comparing Fig. \ref{fig:Optimal_smoothing} (a\&c), it is clear that the pre-emphasis step removes the low-frequency rumbling noise. The contribution of optimal smoothing  in averaging out the effects of late reverberation without destroying the formant structure during speech frames is evident mid and higher frequencies. This is accomplished by using a time-frequency varying optimal smoothing parameter that assigns a higher value in non-speech regions and a lower value in speech regions to the smoothing parameter, as shown in Fig. \ref{fig:Optimal_smoothing} (b). \\

\vspace{0.1em}
\noindent\textbf{[Proposed vs signal processing based algorithms]:} Statistical approaches succeeded in suppressing reverberation, but not to the extent of deep learning approaches. We compare the performance of the proposed network with two signal-processing based approaches WPE, and LPC residual enhancement. Fig. \ref{fig:results} (a\&b) clearly shows that a widely used WPE algorithm does not effectively remove reverberation. This is most likely because this algorithm was designed with the goal of limiting late reflection while preserving early reflection, both of which were discovered to be useful in training robust acoustic models for speech recognition systems \cite{WPE1,WPE2}. Single-channel WPE achieves 14.57\% and 25.64\% relative improvements in PESQ and SNR over unprocessed raw reverberant speech signals. The multi-channel WPE algorithm, as well as multi-channel WPE combined with beamforming, are capable of achieving larger improvements in overall speech quality by utilizing spatial information. However, it was unable to outperform approaches based on deep learning in the majority of speech quality metrics. The proposed SkipConvGAN algorithm outperforms the best performing WPE+beamforming algorithm that utilizes an eight-channel microphone array by 15.38\%, 3.23\% in SNR and PESQ. Similarly, the proposed network improves SNR and PESQ by an absolute of 7.03 dB and 1.189, respectively, over single-channel WPE.
On the other hand, a two-stage dereverberation system based on LPC residual enhancement significantly reduces reverberation. This is evident from the SRMR scores obtained using this approach. This method improves SRMR by 8.11\% relative to eight-channel WPE. However, the proposed approach outperforms the LPC based approach by a relative of 1.02\% in SRMR and outperforms it in all other quality metrics as well.

\vspace{0.1em}
\noindent\textbf{[Proposed vs deep learning based algorithms]:} Next, we compare the proposed network's performance to that of two deep learning-based approaches. The first is a real-valued network that enhances only the magnitude response and reuses the phase from the reverberant speech signal, i.e., SkipConvNet; the second is a complex-valued network that attempts to simultaneously improve both the magnitude and phase, DCCRN. We observe that SkipConvNet was able to efficiently estimate the global structure with defined speech and non-speech boundaries but not the finer formant structure when closely examined at lower and mid frequencies. This strategy outperforms DCRNN in all metrics except PESQ. A possible explanation for this performance is that quality metrics such as cepstral distance, log-likelihood ratio, and SNR are computed primarily from magnitude responses, and SkipConvNet is trained with a loss aimed at reducing the euclidean distance between the magnitude responses of enhanced and clean speech signals. DCRNN outperforms SkipConvNet in PESQ by a relative of 19.68\%. This indicates that enhancing phase is critical for optimizing the perceived quality of reverberant speech signals. Similar to DCCRN, the proposed network is a complex-valued network that is aided during training by a competing discriminator network. Thus, the proposed network is able to outperforms both the SkipConvNet and DCCRN. From Table \ref{tab:Results}, we observe that the proposed network achieves a relative of \{0.43\%, 13.15\%\}, \{10.70\%, 26.97\%\} and \{35.14\%, 12.92\%\} in CD, SNR and PESQ respectively over SkipConvNet and DCCRN. We encourage readers to visit \MYhref{https://vkothapally.github.io/SkipConvGAN/}{https://vkothapally.github.io/SkipConvGAN/} for enhanced speech signals and their corresponding spectrograms that include SkipConvNet, DCCRN, and other approaches not mentioned in this paper.

\vspace{0.1em}
\noindent\textbf{[Proposed vs GAN based algorithms]:} Finally, we compare the performance of the proposed network to that of two other GAN-based approaches. The first network processes reverberant speech signals using magnitude response and is trained with the goal of improving the performance of a back-end speech application, FSEGAN. Later, is a network that is trained on 4-second audio samples with an explicit purpose of improving the speech quality-PESQ, MetricGAN.
FSEGAN performs significantly better than MetricGAN. Although MetricGAN demonstrated promising performance in reducing background noise over a wide range of SNR from -8 dB to 8 dB, its performance in dereverberation was suboptimal in the majority of speech quality metrics, with the exception of PESQ and SRMR. A possible explanation for this performance is that the generator network could have optimized to fool the discrimator, which is biased toward a single metric of speech quality. MetricGAN outperforms the FSEGAN and the proposed GAN by a relative of \{56.03\%, 25.53\%\} in SRMR. However, the proposed GAN outperforms both GAN-based approaches by a relative of \{11.79\%, 39.81\%\}, \{26.64\%, 65.43\%\}, and \{6.67\%, 13.27\%\} in CD, LLR, and PESQ.\\

\vspace{0.1em}
\noindent\textbf{[Summary over averaged speech quality metrics]:} To summarize the results from Tables \ref{tab:Results}, we look at improvements averaged over all speech quality metrics for simulated and real data from the evaluation set. The proposed system increased the SRMR of real recordings of reverberant speech by 3.17 absolute and by 96.63\% relative to unprocessed simulated reverberant speech signals averaged across all speech quality metrics. The proposed network outperformed single-channel WPE and LPC residual enhancement approaches by \{69.15\%, 59.14\} and \{59.75\%, 6.75\%\} respectively on simulated and real recordings. Likewise, the proposed network outperformed deep learning based approaches such as SkipConvNet and DCCRN by a relative of \{15.59\%, 22.64\%\} on simulated recordings. However, SkipConvNet outperforms both DCCRN and the proposed network by a relative of \{39.11\%, 11.09\%\} on real recordings. Furthermore, the proposed network outperforms other GAN-based approaches for dereverberation of single-channel speech. For example, the proposed network outperformed FSEGAN by 16.29\% and 9.89\% on simulated and real recordings respectively.

\section{Conclusion}
In naturalistic environments, reverberation causes one of the most significant speech distortions in speech captured by distant microphones. In addition, background noise such as air conditioning units, computer fans, etc., reduces speech intelligibility and quality. In this study, we present `SkipConvGAN' as a novel complex-valued generative adversarial network with the following key contributions: (i) the proposed network employs fully convolutional complex-valued networks for both generator and discriminator networks, (ii) to bridge the semantic gap between features from the encoder and decoder layers, the proposed network employs complex-valued skipconv blocks within the generator network's skip connections, (iii) we use a complex-valued time-frequency self-attention (TF-SA) module that attends to features in both time and frequency dimensions while also preserving the interdependence of the real and imaginary components of the intermediate feature maps, and (iv) the proposed network uses a patch-discriminator and is trained using a combination of adversarial and feature loss. Furthermore, we employ a time-frequency varying optimal smoothing strategy as a pre-processing step to reduce late reflections in training samples. First, we conduct an ablation study to assess the contribution of the aforementioned components in the generator network of the proposed system towards overall speech quality improvements. From evaluations, we conclude that the proposed components have progressively enhanced speech quality. Next, we compare the proposed network's performance to that of signal processing and deep learning approaches. We used the REVERB challenge corpus, which included simulated and real recordings with varying levels of reverberation. Evaluations show that, when compared to other algorithms, the proposed network is more effective at restoring lost formant structure and thus improving overall speech quality. Despite the fact that the proposed network was designed for the dereverberation task, we found it to be resistant to moderate levels of non-speech background noise in the speech. We intend to extend the proposed network in the future to address joint speech-denoising and speech-dereverberation using multi-channel distant speech capture.

\vspace{-1em}

\ifCLASSOPTIONcaptionsoff
  \newpage
\fi

\bibliographystyle{IEEEtran}
\bibliography{mybib}

\begin{IEEEbiography}[{\includegraphics[width=1.0in,height=1.25in,clip]{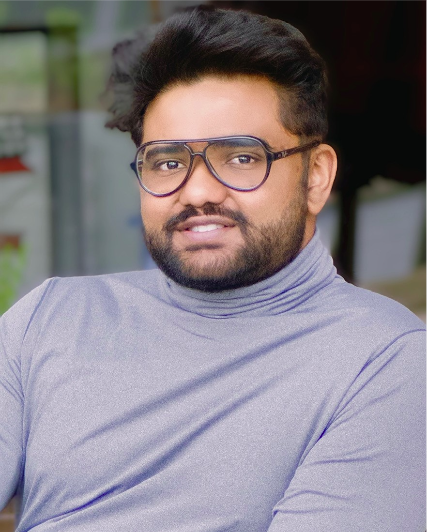}}]%
	{Vinay Kothapally}
	
	is a graduate research assistant at the University of Texas at Dallas' Center for Robust Speech Systems (CRSS) since the fall of 2016. He earned his Master of Science in Electrical Engineering from Missouri University of Science and Technology (MS\&T) in Rolla,MO, before beginning his Ph.D. program in Electrical Engineering at UT Dallas. He worked on spatial audio as a DSP Audio Intern at GoPro in San Mateo, CA. He also interned at CypherCorp and AuSIM as an audio intern. He worked as a research intern at Facebook Reality Labs in Seattle in the fall of 2019. His research at CRSS focuses on the development of single-channel and multi-channel solutions for improving the robustness of speech systems against distortions caused by distant speech capture. His other areas of interest include back-end speech applications, such as speech and speaker recognition.
\end{IEEEbiography}	
\begin{IEEEbiography}[{\includegraphics[width=1.0in,height=1.25in,clip]{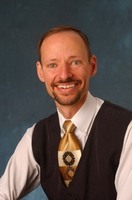}}]%
{John H. L. Hansen}
(S'81--M'82--SM'93--F'07) received the Ph.D. and M.S. degrees in electrical engineering from Georgia Institute of Technology, Atlanta, GA, USA, in 1988 and 1983, respectively,and the B.S.E.E. degree from Rutgers University, College of Engineering, New Brunswick, NJ, USA, in 1982. He received the honorary degree Doctor Technices Honoris Causa from Aalborg University, Aalborg, Denmark, in April 2016 in recognition of his contributions to speech signal processing and speech/language/hearing sciences. He joined the University of Texas at Dallas (UT Dallas), Erik Jonsson School of Engineering and Computer Science, Richardson, TX, USA in 2005, where he currently serves as Jonsson School Associate Dean for Research, as well as a Professor of electrical and computer engineering, the Distinguished University Chair in Telecommunications Engineering, and a joint appointment as a Professor in the School of Behavioral and Brain Sciences (Speech \& Hearing). He previously served as the Department Head of Electrical Engineering from August 2005-December 2012, overseeing a +4x increase in research expenditures (4.5 M-22.3 M) with a 20\% increase in enrollment along with hiring 18 additional T/TT faculty, growing UT Dallas to the 8th largest EE program from ASEE rankings in terms of degrees awarded. At UT Dallas, he established the Center for Robust Speech Systems (CRSS). Previously, he served as the Department Chairman and a Professor of the Department of Speech, Language and Hearing Sciences (SLHS), and a Professor in electrical \& computer engineering, University of Colorado \-Boulder (1998-2005), where he co-founded and served as the Associate Director of the Center for Spoken Language Research. In 1988, he established the Robust Speech Processing Laboratory and continues to direct research activities in CRSS, UT Dallas. He is author/coauthor of 661 journal and conference papers including 12 textbooks in the field of speech processing and language technology, signal processing for vehicle systems, coauthor of textbook Discrete-Time Processing of Speech Signals, (IEEE Press, 2000), co-editor of DSP for InVehicle and Mobile Systems (Springer, 2004), Advances for In-Vehicle and Mobile Systems: Challenges for International Standards (Springer, 2006), InVehicle Corpus and Signal Processing for Driver Behavior (Springer, 2008), and lead author of the report The Impact of Speech Under Stress on Military Speech Technology, (NATO RTO-TR-10, 2000). His research interests include the areas of digital speech processing, analysis and modeling of speech and speaker traits, speech enhancement, feature estimation in noise, robust speech recognition with emphasis on spoken document retrieval, and in-vehicle interactive systems for hands-free human-computer interaction. He has been named IEEE Fellow (2007) for contributions in “Robust Speech Recognition in Stress and Noise,” International Speech Communication Association (ISCA) Fellow (2010) for contributions on research for speech processing of signals under adverse conditions, and received The Acoustical Society of Americas 25 Year Award (2010) in recognition of his service, contributions, and membership to the Acoustical Society of America. He is currently serving as the ISCA President (2017-19) and a member of the ISCA Board, having previously served as the Vice-President (2015-17). He also was selected and is serving as the Vice-Chair on U.S. Office of Scientific Advisory Committees (OSAC) for OSAC-Speaker in the voice forensics domain (2015-2017). Previously, he served as the IEEE Technical Committee (TC) Chair and Member of the IEEE Signal Processing Society: Speech-Language Processing Technical Committee (SLTC) (2005-08; 2010-14; elected IEEE SLTC Chairman for 2011-13, Past-Chair for 2014), and elected ISCA Distinguished Lecturer (2011-12). He has served as member of IEEE Signal Processing Society Educational Technical Committee (2005-08; 2008-10); Technical Advisor to the U.S. Delegate for NATO (IST/TG-01); IEEE Signal Processing Society Distinguished Lecturer (2005/06), an Associate Editor of the IEEE TRANSACTIONS ON SPEECH \& AUDIO PROCESSING (1992-99), an Associate Editor of the IEEE SIGNAL PROCESSING LETTERS (1998-2000), Editorial Board Member of the IEEE Signal Processing Magazine (2001-03); and a Guest Editor (October 1994) for special issue on Robust Speech Recognition of IEEE TRANSACTIONS ON SPEECH \& AUDIO PROCESSING. He has served on Speech Communications Technical Committee for Acoustical Society of America (2000-03), and previously on ISCA Advisory Council. He has supervised 82 Ph.D./M.S. thesis candidates (45 Ph.D., 37 M.S./M.A.), received The 2005 University of Colorado Teacher Recognition Award as voted on by the student body. He also organized and served as the General Chair for ISCA Interspeech- 2002, September 16-20, 2002, Co-Organizer and Technical Program Chair for IEEE ICASSP-2010, Dallas, TX, March 15-19, 2010, and Cochair and Organizer for IEEE SLT-2014, December 7-10, 2014 in Lake Tahoe, NV, USA.
\end{IEEEbiography}

\end{document}